\documentclass{article}
\usepackage{amsfonts}
\usepackage{amsmath}

\input{tcilatex}

\begin{document}

\title{More General Sudden Singularities }
\author{John D. Barrow \\
DAMTP, Centre for Mathematical Sciences,\\
Cambridge University, \\
Wilberforce Road,\\
Cambridge CB3 0WA, UK}
\date{}
\maketitle

\begin{abstract}
We present a general form for the solution of an expanding
general-relativistic Friedmann universe that encounters a singularity at
finite future time. The singularity occurs in the material pressure and
acceleration whilst the scale factor, expansion rate and material density
remain finite and the strong energy condition holds. We also show that the
same phenomenon occurs, but under different conditions, for Friedmann
universes in gravity theories arising from the variation of an action that
is an arbitrary analytic function of the scalar curvature.
\end{abstract}

There has been considerable interest in the possibility of finite and
infinite future-time cosmological singularities in general relativity under
the headings of 'big-rip' singularities and associated 'phantom matter'
evolution \cite{rip}\cite{diaz}. These studies have also extended to several
quantum and brane-world generalisations \cite{calcag}, \cite{pasq}\cite%
{kofinas}, \cite{sahni}, \cite{noj1,noj2}. A demarcation of these and
related finite-time singularities has been given in \cite{cot}.

We consider the most general form of sudden, or finite-time, singularity of
the sort discussed in \cite{jb} that can arise in an Friedmann expanding
universe. In \cite{jb} we showed that singularities of this sort can arise
from a divergence of the pressure, $p$, at finite time despite the scale
factor, $a(t)$, the material density, $\rho ,$ and the Hubble expansion
rate, $H=\dot{a}/a$ remaining finite. The pressure singularity is
accompanied by a divergence in the acceleration of the universe, $\ddot{a}$,
at finite time. Remarkably, these singularities occur without violating the
strong energy conditions $\rho >0$ and $\rho +3p>0.$ They can even prevent a
closed Friedmann universe that obeys these energy conditions from attaining
an expansion maximum \cite{jb2}. In order to prevent the occurrence of these
sudden singularities it is necessary to bound the pressure. A sufficient
condition is to require $dp/d\rho $ to be continuous or $p/\rho $ to be
finite \cite{jb}.

Consider the Friedmann universe with curvature parameter $k$, and $8\pi
G=c=1 $, the Einstein equations reduce to

\begin{eqnarray}
3H^{2} &=&\rho -\frac{k}{a^{2}},  \label{fried} \\
0 &=&\dot{\rho}+3H(\rho +p),  \label{con} \\
\frac{\ddot{a}}{a} &=&-\left( \frac{\rho +3p}{6}\right) .  \label{fr}
\end{eqnarray}%
In \cite{jb} we constructed an explicit example by seeking, over the time
interval $0<t<t_{s},$ a solution for the scale factor $a(t)$ of the form

\begin{equation}
a(t)=1+Bt^{q}+C(t_{s}-t)^{n},  \label{ex}
\end{equation}%
where $B>0,q>0,C$ and $n>0$ are free constants to be determined. If we fix
the zero of time by requiring $a(0)=0,$ so $Ct_{s}^{n}=-1,$ we have

\begin{equation}
a(t)=(\frac{t}{t_{s}})^{q}\left( a_{s}-1\right) +1-(1-\frac{t}{t_{s}})^{n},
\label{sol2}
\end{equation}%
where $a_{s}\equiv a(t_{s})$. Hence, as $t\rightarrow t_{s}$ from below, we
have

\begin{equation}
\ddot{a}\rightarrow q(q-1)Bt^{q-2}-\frac{\ n(n-1)}{t_{s}^{2}(1-\frac{t}{t_{s}%
})^{2-n}}\rightarrow -\infty  \label{Lim}
\end{equation}%
whenever $1<n<2$ and $0<q\leq 1$; the solution exists on the interval $%
0<t<t_{s}$. Hence, as $t\rightarrow t_{s\text{ }}$we have $a\rightarrow
a_{s} $, $H\rightarrow H_{s}$ and $\rho \rightarrow \rho _{s}>0$ where $%
a_{s},H_{s},$ and $\rho _{s}$ are all finite but $p_{s}\rightarrow \infty $.
By contrast, as $t\rightarrow 0$ there is an initial all-encompassing
strong-curvature singularity, with $H\rightarrow \infty ,\rho \rightarrow
\infty $ and $p\rightarrow \infty $. From (\ref{Lim}) and (\ref{fr}), we see
that $\rho $ and $\rho +3p$ remain positive throughout the evolution but,
because $\rho $ is finite asymptotically, the dominant-energy condition, $%
\left\vert p\right\vert \leq \rho ,$ must always be violated as $%
t\rightarrow t_{s}$, \cite{jb}. A solution with similar properties which
expands from a de Sitter past state also exists, with

\begin{equation}
a(t)=a_{s}-1+\exp \{\lambda (t_{s}-t)\}-(1-\frac{t}{t_{s}})^{n}  \label{exp}
\end{equation}%
with $\lambda >0$ constant and $1<n<2$.

The particular solution (\ref{sol2}) expands with

\begin{equation}
a(t)\approx n\frac{t}{t_{s}}+\left( \frac{t}{t_{s}}\right)
^{q}(a_{s}-1)\approx \left( \frac{t}{t_{s}}\right) ^{q}  \label{zero}
\end{equation}%
as $t\rightarrow 0$ and so with choices $q=1/3,1/2,$ $2/3,$or $q=1$ it will
resemble a massless scalar-, radiation-, dust-, or
negative-curvature-dominated Friedmann universe respectively at early times.
At late times, as the 'big-rip' singularity is approached, the expansion
approaches a constant with

\begin{equation}
a(t)\approx a_{s}+q(1-a_{s})(1-\frac{t}{t_{s}})  \label{asy}
\end{equation}%
as $t\rightarrow t_{s}$.

Consider now the most general form of a solution of this type. In the
neighbourhood of the singularity at $t=t_{s}$ the general formal solution of
the Friedmann equations has the form

\begin{equation}
a(t)=(t_{s}-t)^{n}\left[ \sum_{j=0}^{\infty
}\sum_{k=0}^{N_{j}}a_{jk}(t_{s}-t)^{j/Q}(\log ^{k}[t_{s}-t])\right]
\label{psi}
\end{equation}%
where $[...]$ is a convergent double Psi-series which tends to zero as $%
t\rightarrow t_{s}$; $a_{jk}$ are constants, $N_{j}\leq j$ are positive
integers and $Q\in 
\mathbb{Q}
^{+}$ (this is derived as an application of the theorem of Goriely and Hyde
proved in ref. \cite{gor})$.$

Hence, we have a generalisation of the solution (\ref{sol2}) of the form

\begin{equation}
a(t)=(\frac{t}{t_{s}})^{q}\left( a_{s}-1\right) +1-(t_{s}-t)^{n}\left[
\sum_{j=0}^{\infty }\sum_{k=0}^{N_{j}}a_{jk}(t_{s}-t)^{j/Q}(\log
^{k}[t_{s}-t])\right] .  \label{gen}
\end{equation}%
As $t\rightarrow t_{s}$ if we choose $1<n<2$, $0<q<1$, we have $a\rightarrow
a_{s}$ and

\begin{equation}
\ddot{a}\rightarrow -n(n-1)A(t_{s}-t)^{n-2}\log ^{k}(t_{s}-t)+\text{higher
order}\rightarrow -\infty ,  \label{genlim}
\end{equation}%
where $A>0$ is constant,and so $\rho +3p>0$ and $\ p>\left\vert \rho
\right\vert $ as the future singularity is approached with $p\rightarrow
\infty $ at $t_{s}$. As $t\rightarrow 0$ the solution evolves as $a\approx
c_{0}+c_{1}t^{q}$ and approaches the behaviour of the radiation-dominated
Friedmann universe for $q=1/2$. As $t\rightarrow \infty $ the solution
approaches $a_{s}$ in accord with eq.(\ref{asy}). We see that the solution (%
\ref{sol2}) is the special case with all $a_{jk}=0$ except for $a_{00}\neq 0$%
.

By choosing $n$ to lie in the intervals $(N,N+1)$ for $N\geq 2$ where $N\in 
\mathbb{Z}
^{+},$ we can create a singularity in which

\begin{equation}
\frac{d^{N+1}a}{dt^{N+1}}\rightarrow \infty  \label{diff}
\end{equation}%
but

\begin{equation}
\frac{d^{r}a}{dt^{r}}\rightarrow 0,\text{ }for\text{ }r\leq N\in 
\mathbb{Z}
^{+}.  \label{diff2}
\end{equation}

In particular, this allows for singularities of the pressure which are
accompanied by divergences of higher time-derivatives of $a(t)$ in Friedmann
solutions of higher-order gravity theories \cite{jb3}. Consider a
generalisation of general relativity to a theory of gravity derived from the
variation of a lagrangian that is an arbitrary polynomial function $f(R)$ of
the scalar curvature, $R$. The Friedmann universe is now prescribed by the
solution of any two of equations (\ref{con}) and the generalisations of (\ref%
{fr}) and (\ref{fried}), which are given respectively by \cite{jb3}

\begin{equation}
Rf^{\prime }-2f+3f^{\prime \prime }(\ddot{R}+3\frac{\dot{a}}{a}\dot{R}%
)+3f^{\prime \prime \prime }\dot{R}^{2}=3p-\rho  \label{f4}
\end{equation}%
and

\begin{equation}
3\frac{\ddot{a}}{a}f^{\prime }+\frac{1}{2}f-3f^{\prime \prime }\dot{R}\frac{%
\dot{a}}{a}+\rho =0  \label{f3}
\end{equation}%
where $f^{\prime }\equiv df/dR$ etc, and the scalar curvature of the
Friedmann metric is given by

\begin{equation}
R=-6\left( \frac{\ddot{a}}{a}+\frac{\dot{a}^{2}}{a^{2}}+\frac{k}{a^{2}}%
\right) .  \label{R}
\end{equation}%
The equations of general relativistic cosmology in the absence of a
cosmological constant arise from (\ref{f4})-(\ref{R}) as the special case $%
f(R)=R.$ Various auxiliary conditions on the form of $f(R)$ are required to
ensure the existence of Friedmann solutions, see ref. \cite{jb3} for a
discussion. We ignore them as they do not affect our results here.

In order to obtain a finite-time singularity of the sort described by (\ref%
{sol2}) or (\ref{gen}) we need to accommodate a singularity of the pressure, 
$p$, in (\ref{con}) in equations (\ref{f4})-(\ref{f3}). We require that all
the terms in the generalised Friedmann first integral (\ref{f3}) remain
finite as $t\rightarrow t_{s}$ and $a\rightarrow a_{s}$. Thus we require
that $\ddot{a},f(R),f^{\prime }(R),f^{\prime \prime }(R),\dot{R},R$ and $%
\rho $ are all finite in this limit.

Now consider the evolution equation (\ref{f4}) as $t\rightarrow t_{s}.$
Since $R,f,f^{\prime },f^{\prime \prime }$ and $f^{\prime \prime \prime }$
will be finite at $t_{s}$ we must have

\begin{equation}
f^{\prime \prime }\ddot{R}\rightarrow p\text{ }\rightarrow -\infty \text{ as 
}t\rightarrow t_{s}.  \label{div}
\end{equation}%
Hence, there must be a divergence of

\begin{equation}
\ddot{R}\rightarrow -6\frac{\ddddot{a}}{a}\rightarrow -\infty \text{ as }%
t\rightarrow t_{s}  \label{h}
\end{equation}%
but with $R(t_{s})$ and $\dot{R}(t_{s})$ remaining finite. This occurs so
long as $3<n<4,$ since for a solution of the form (\ref{sol2})

\begin{equation}
\lim_{t\rightarrow t_{s}}\frac{d^{r}a}{dt^{r}}=-\frac{n(n-1)..(n-r+1)}{%
t_{s}^{r}}(1-\frac{t}{t_{s}})^{n-r}  \label{lim}
\end{equation}%
and the same condition holds for the general solution (\ref{gen}). In the
general relativity case of $f(R)=R,$ the pressure singularity accompanies
one in $\ddot{a}$ and requires $1<n<2.$

This generalisation of the conditions for the existence of sudden
singularities to the Friedmann universes of higher-order lagrangian theories
of gravity shows that the higher-order curvature corrections that these
extensions introduce are able to moderate but not remove the finite-time
singularity. They merely change the evolutionary behaviours of the expansion
scale factor which permit their occurrence. As in \ general relativity, an
energy condition that bounds $p/\rho $ suffices to exclude them.

\end{document}